**A Proton Treatment Planning Method for Combining FLASH and Spatially Fractionated Radiation Therapy to Enhance Normal Tissue Protection**

Weijie Zhang, Xue Hong, Ya-Nan Zhu, Yuting Lin, Gregory Gan, Ronald C Chen, and Hao Gao

Department of Radiation Oncology, University of Kansas Medical Center, USA

**Email:** wzhang2@kumc.edu, hgao2@kumc.edu

**Abstract**

**Background:** FLASH radiation therapy (FLASH-RT) enhances normal tissue sparing through the ultra-high dose rate irradiation known as the FLASH effect. In proton Bragg peak FLASH-RT this effect is generally limited to high-dose regions in normal tissue close to the target (deep tissue). Spatially Fractionated Radiation Therapy (SFRT) creates distinct spatial dose distributions, consisting of alternating high-dose ("peak") and low-dose regions ("valley"), to activate various biological mechanisms that improve normal tissue protection, characterized by the peak-to-valley dose ratio (PVDR). Due to proton's multiple Coulomb scattering, the biological sparing effect of SFRT with high PVDR is primarily seen in normal tissues from the beam entrance to shallow-to-intermediate depths, a few centimeters from the target. With the therapeutic potential of each technique established independently, the combination of FLASH-RT and SFRT could offer a powerful and synergistic approach for improved treatment outcomes.

**Purpose:** In this work, the treatment planning study is performed to investigate the possibility of a new proton modality SFRT-FLASH that synergizes FLASH-RT and SFRT for enhanced normal tissue protection, i.e., the use of FLASH-RT to enhance the sparing of deep-depth normal tissues and the use of SFRT to enhance the sparing of shallow-to-intermediate-depth normal tissues.

**Methods:** Two SFRT techniques, proton GRID therapy with conventional beam size (pGRID) and proton minibeam radiation therapy (pMBRT), are considered for SFRT-FLASH, i.e., pGRID-FLASH (SB-FLASH) and minibeam-FLASH (MB-FLASH). pGRID utilizes the scissor-beam (SB) method to achieve uniform dose distribution in target. To achieve the FLASH effect's high-dose (5 Gy) and high-dose-rate (40 Gy/s) thresholds, a single-field-uniform-dose-per-fraction (SFUDPF) delivery strategy is employed. In addition to conventional dose constraints, a dose rate constraint is applied to CTV1cm (an auxiliary organ-at-risk (OAR) structure defined as a 1 cm ring extension of the CTV, excluding the CTV) for each field. The dose and dose rate objectives are jointly optimized during treatment planning.

**Results:** The proposed methods (MB-FLASH and SB-FLASH) were validated in comparison to conventional (CONV), FLASH-RT (FLASH), pMBRT (MB), and pGRID (SB) plans across four clinical cases. Our method achieved both high FLASH effect coverage near the target and high PVDR in shallow-to-intermediate depths. For example, the CTV1cm volume achieved ~60-80% FLASH effect coverage, and intermediate-depth dose planes in the beam-eye-view achieved PVDR values of approximately 2.5-7.

**Conclusions:** We present a novel proton treatment planning approach that achieves the FLASH effect at deep tissue depths while maintaining high PVDR at shallow-to-intermediate depths, enhancing normal tissue protection and advancing the therapeutic potential of proton therapy.

# 1. Introduction

Normal tissue toxicity remains a primary factor limiting the effectiveness of radiation therapy (RT) [1,2]. As such, there is increasing interest in modalities that can offer enhanced protection to normal tissue. Among these, spatially fractionated radiotherapy (SFRT) [3-10] and FLASH radiotherapy (FLASH-RT) [11-17] stand out as promising approaches. The unique dose delivery characteristics of FLASH-RT and SFRT offer additional biological normal tissue sparing compared to conventional RT (CONV) [4,14,18-20].

FLASH-RT focuses on the temporal aspect of dose delivery, delivering ultra-high dose rates ($\geq 40$ Gy/s) and high doses ($\geq 5$ Gy) to achieve what is known as the FLASH effect. In vitro studies have shown that FLASH tissue-sparing and cellular oxygen depletion starts at 40 Gy/s and 5Gy [21]. This effect has been shown to significantly reduce normal tissue toxicity while maintaining anti-tumor efficacy [22,23]. In contrast, SFRT emphasizes spatial modulation of the dose, creating high-dose "peak" regions interspersed with low-dose "valley" regions. There are four main types of SFRT, i.e., GRID [24], LATTICE [25], minibeams [26], and microbeams [27].

Proton RT is advantageous in dose shaping for its selective depth dose deposition and negligible dose deposition after the Bragg peak compared to photon RT. Researchers have explored the delivery of FLASH-RT or SFRT using proton RT to capitalize on these advantages [16,22,28-36].

This study aims to explore a combined approach using proton SFRT and FLASH-RT, i.e., SFRT-FLASH, to create treatment plans with enhanced normal tissue protection [37]. The FLASH effect usually only occurs in high-dose normal tissue region close to the treatment target (i.e., deep depth). We consider two different modalities of SFRT, proton GRID (pGRID) and proton minibeam RT (pMBRT), i.e., pGRID-FLASH (SB-FLASH) and pMBRT-FLASH (MB-FLASH) respectively for SFRT-FLASH. Due to multiple coulomb scattering of protons, pGRID and pMBRT plans can have homogeneous dose in the deep depth and heterogeneous dose in the shallow-to-intermediate depth, where high peak-to-valley dose ratio (PVDR) can provide better normal tissue sparing. Therefore, SFRT-FLASH has the potential to offer superior normal tissue sparing by leveraging the unique benefits of each technique.

This work for SFRT-FLASH only considers uniform prescribed target dose, to align with current clinical standard of CONV, which may lead to broad applications of SFRT-FLASH. pGRID utilizes the scissor-beam (SB) method [38] to achieve uniform dose distribution in target. To meet the high-dose threshold and achieve optimal dose rate and robustness, we delivered the target dose through a single field per fraction using a hypofractionated regimen, such as an 8-Gy single beam per fraction for five fractions for MB-FLASH or a pair of 8-Gy scissor beams [38] per fraction for two fractions during SB-FLASH. This delivery strategy is termed single-field-uniform-dose-per-fraction (SFUDPF). The pencil beam scanning (PBS) dose rate model is employed [39] to evaluate the FLASH effect. SFRT-FLASH will be evaluated in comparison with CONV, FLASH-RT, and SFRT, demonstrating SFRT-FLASH can synergize SFRT and FLASH, to achieve sufficient FLASH effect coverage near the target and at the same time high PVDR in shallow-to-intermediate depths for enhanced normal tissue protection.

**2. Methods and Materials**

*2.1. SFRT techniques*

This study evaluates two SFRT techniques for SFRT-FLASH: pGRID and pMBRT. For both techniques, the goal is to deliver a uniform dose to the target to maximize anti-tumor efficacy and fully utilize the advantages of proton RT.

To avoid the use of physical collimators and beam blocking, we employ the SB method [38] to generate a GRID dose pattern at varying depths. The SB technique involves adding a complementary beam, tailored to the depth of the organ-at-risk (OAR) of interest, to the primary beam. This approach allows two beams to maintain uniform dose delivery to the target while optimizing the spatially fractionated dose distribution in normal tissues without the need for a physical collimator.

For pMBRT, a multi-slit collimator is used to produce a submillimeter minibeam dose pattern. The collimator parameters follow those outlined in [35], with a slit width of 0.4 mm and a center-to-center (ctc) distance varying from 2 mm to 7 mm, depending on the tumor depth. By selecting different ctc distances for each fraction, we can balance target dose uniformity with spatially fractionated dose distribution in normal tissues, achieving a high-quality treatment plan.

*2.2. SFUDPF delivery and optimization*

Given that the dose threshold for the FLASH effect is 5 Gy, the prescribed target dose in this study is set using a hypofractionated regimen of 8 Gy (a single beam) per fraction over 5 fractions for pMBRT and 16 Gy (a pair of SB with 8Gy per beam) per fraction over 2 fractions for pGRID. The FLASH effect is determined per beam and deemed to be effective only if both dose and dose rate thresholds are met per beam, independent of other beams. To align with current clinical standard of CONV, we consider a uniform dose delivery to the target for each fraction. This delivery approach is referred to as SFUDPF. In the case of pGRID with the SB setup, each field includes both a primary and a complementary beam to ensure dose uniformity across each fraction.

We start by defining the conventional Intensity-Modulated Proton Therapy (IMPT) optimization problem for baseline comparison and notation purposes:

$$\min_x f(d)$$
$$\text{s.t.} \begin{cases} d = Ax \\ x \in \{0\} \cup [G, +\infty] \end{cases} \quad (1)$$

In Eq. (1), $x$ represent proton spot weights to be optimized, $d$ denotes the 3D dose distribution, and $A$ is the dose influence matrix. Function $f(d)$ represents the planning objective function for conventional IMPT. To ensure deliverability, the minimum monitor unit (MMU) constraint is enforced, where $x$ must be either zero or meet the MMU threshold G. For conventional IMPT, G is generally set to 5. It's important to note that the conventional IMPT plan maintains consistent spot weights $x$ across fractions, so each fraction has identical weights.

In contrast to the conventional IMPT optimization in Eq. (1), the spot weights $x$ of SFUDPF vary across fractions. The SFUDPF objective function incorporates additional regularization terms to control dose distribution to the target for each fraction. We present the SFUDPF optimization problem as follows:

$$\min_x f(d) + \sum_{k=1}^{K} g(d_k)$$
$$\text{s.t.} \begin{cases} x = [x_1, x_2, ..., x_K] \\ d = d_1 + d_2 + ... + d_K \\ d_k = A_k x_k, k = 1, ..., K \\ x_k \in \{0\} \cup [G, +\infty], k = 1, ..., K \end{cases} \quad (2)$$

In Eq. (2), *K* represents the total number of fractions, the spot weights *x* to be optimized include the spot weights for all fractions. The cumulative dose distribution *d* is obtained by summing the dose contributions from each fraction, with $A_k$ denoting the dose influence matrix specific to the *k*-th fraction. In the SFUDPF optimization problem, in addition to the conventional objective function *f(d)*, introduce additional target control terms $g(d_k)$ for each fraction *k*. The objective *f(d)* includes standard target and OAR-specific planning goals based on dose-volume histogram (DVH) constraints for the total dose. In contrast, the additional term $g(d_k)$ applies only to the target for each individual fraction *k*, ensuring that the dose distribution in each fraction meets desired uniformity and dose constraints independently.

*2.3. Dose rate optimization*

To achieve the FLASH effect in high-dose region near the target, an additional requirement of ultra-high dose rates must be met, which is typically not achievable with conventional IMPT plans. In this work, in addition to the dose optimization in Eq. (2), we incorporate the strategy described in [40] to maximize the tissue volume receiving the desirable FLASH dose rate or higher. The optimization problem of FLASH plans is given as follows

$$\min_x f(d) + \sum_{k=1}^{K} g(d_k)$$
$$s.t. \begin{cases} x = [x_1, x_2, ..., x_K] \\ d = d_1 + d_2 + ... + d_K \\ d_k = A_k x_k, k = 1, ..., K \\ x_k \in \{0\} \cup [G, +\infty], k = 1, ..., K \\ V_{\gamma_k \geq \gamma_0, ROI} \geq P\% \end{cases} \quad (3)$$

In Eq. (3), $\gamma_k$ denotes the dose rate distribution for the $k$-th fraction. The region of interest (ROI), where the dose rate constraint is applied, is defined as CTV1cm, a 1 cm ring extension surrounding the CTV. A straightforward constraint is imposed: at least P% of the voxels within the ROI must achieve a dose rate of at least $\gamma_0=40$ Gy/s.

We calculated the dose rates using the PBS dose rate model [39]. In this model, the dose received of a voxel v in a single fraction is a time-dependent function $D'(v,t)$. The full fraction delivery time, $t_f$, represents the total dose delivered, $D(v)=D'(v,t_f)$. The delivery time $T(v)$ of a voxel $v$ is defined as the duration from receiving 2.5% to 97.5% of its total dose $D(v)$. The dose rate of this voxel is then calculated as 95% of its dose $D(v)$ divided by the delivery time $T(v)$:

$$\begin{aligned}
D'(v,T_0) &= 0.025 D(v) \\
D'(v,T_1) &= 0.975 D(v) \\
T(v) &= T_1 - T_0 \\
\gamma(v) &= \frac{0.95 D(v)}{T(v)}
\end{aligned} \quad (4)$$

*2.4. Solution algorithm*

We have established our optimization problem for the SFRT-FLASH techniques as shown in Eqs. (3) and (4). Here, we describe the algorithm used to solve this problem. The terms *f(d)* and *g(d_k)*, which include general DVH based planning constraints, are handled using the iterative convex relaxation (ICR) method and Alternating Direction Method of Multipliers (ADMM) [40-48]. The MMU constraints can be addressed through post-processing and MMU optimization techniques [49-55]. For the PBS dose rate constraints, we applied the strategy described in [56]. Our algorithm works in the following steps

1. Initialization for spot weights $x=0$
2. For outer loop
3.    Calculate the delivery time T for all voxels in ROI and all fractions according to Eq. (4)
4.    Calculate the dose distribution using dose influence matrices, and the dose rate in ROI using the delivery time T in Step 3 for all fractions
5.    Employ ICR method to translate all DVH constraints and dose rate constraint to L2 constraints
6.    For inner loop
7.       Employ ADMM method to solve the problem in Step 5 with MMU constraint
8.    End inner loop
9.    Update spot weights $x$ with post processing for MMU constraint
10. End outer loop

*2.5. Materials*

Two different modalities are evaluated in this study: pGRID-FLASH ("SB-FLASH") and pMBRT-FLASH ("MB-FLASH"). These modalities are compared with conventional IMPT ("CONV"), FLASH with dose rate optimization ("FLASH"), pMBRT ("MB"), pGRID ("SB") acoss head-and-neck (HN), abdomen, lung, and prostate cases. Clinically routine beam angles were chosen: (45º, 135º, 225º,315º) for HN, (45º, 135º, 225º, 315º) for abdomen, (0º, 120º, 240º) for lung, and (90º, 180º) for prostate. All plans are treated with single-field uniform dose delivery, using 8 Gy per fraction over 5 fractions for pMBRT (HN, abdomen and lung) and 16 Gy per fraction over 2 fractions for pGRID (prostate). All plans were normalized to have $D_{95}$=100% to the CTV for each single-field plan.

Beam data was obtained from [57] and dose influence matrices were generated using MatRad [57] with a 5 mm lateral spacing on $1\times1\times3$ mm$^3$ dose grid for pMBRT plans (MB and MB-FLASH) and $3\times3\times3$ mm$^3$ dose grid for other plans. For pMBRT plans, the collimator parameters were set with a 0.4 mm open slit size and ctc distances as follows: (3, 3, 3, 3) mm on (45º, 135º, 225º,315º) for HN, (2, 3, 4, 4) mm on (45º, 135º, 225º,315º) for abdomen, (2, 5, 7) mm on (0º, 120º, 240º) for lung.

In the tables, the conformal index (CI) is defined as $V_{100}^2/(V\times V´_{100})$ ($V_{100}$: CTV volume receiving at least 100% of prescription dose; V: CTV volume; V´$_{100}$: the total volume within the body receiving at least 100% of the prescription dose; for MB and MB-FLASH plans, we consider only the CTV and a 3 cm extension around the CTV to exclude peak doses at shallow depths.; ideally CI=1); We adopt the formula in [58] to calculate PVDR. For each beam angle, we select a depth that we interested in and use the dose on the beam-eye-view (BEV) plan at this depth to calculate PVDR, i.e., $PVDR=D_{peak}/D_{valley}$, where $D_{peak}$ is the mean dose of all the peak doses and $D_{valley}$ is the mean dose of all the valley doses; $P_{FLASH}$ is the percentage of the ROI volume receiving at least 5 Gy dose and 40 Gy/s dose rate, which is calculated for each single field plan respectively.

## 3. Results

The results for HN, abdomen, lung, prostate are presented in Figures 1-4 respectively. The plan parameters (dose parameters, PVDR and $P_{FLASH}$) are summarized in Tables 1-4 respectively.

*3.1. PVDR*

The comparison of PVDR values between CONV, FLASH, MB and MB-FLASH in Tables 1-3 indicates that MB and MB-FLASH improved the PVDR from CONV and FLASH significantly, i.e., an increase from CONV (1.41, 1.47, 1.31, 1.11) and FLASH (1.22, 1.43, 1.58, 1.15) to MB (2.52, 2.31, 2.76, 2.77) and MB-FLASH (2.76, 2.48, 2.88, 2.84) for (45º, 135º, 225º,315º) in HN, from CONV (1.2, 1.1, 1.0, 1.0) and FLASH (1.2, 1.1, 1.0, 1.1) to MB (17.8, 2.6, 4.4, 3.4) and MB-FLASH (10.2, 2.6, 4.4, 3.5) for (45º, 135º, 225º,315º) in abdomen, from CONV (1.3, 1.1, 1.1) and FLASH (1.4 ,1.2, 1.1) to MB (2.2, 3.3, 6.1) and MB-FLASH (2.2, 3.4, 6.1) for (0º, 120º, 240º) in lung. Moreover, we could find that MB-FLASH has similar or better PVDR compared to MB in most fields except 45º in abdomen, which means that high MMU constraints and high dose rate constraints won't affect the minibeam dose pattern and normal tissue sparing at shallow depth. The difference between peak and valley dose is also clear from the figures, i.e., MB (Figures 1d, 2d, 3d) and MB-FLASH (Figures 1e, 2e, 3e) compared to CONV (Figures 1a, 2a, 3a) and FLASH (Figures 1b, 2b, 3b). The improved PVDR by MB and MB-FLASH from CONV and FLASH is also demonstrated by BEV dose plane plots in Figures 1g, 2g, 3g and dose profile plots in Figures 1h, 2h, 3h.

The comparison of PVDR values between CONV, FLASH, SB and SB-FLASH in Table 4 indicates that SB and SB-FLASH improved the PVDR from CONV and FLASH significantly, i.e., an increase from CONV (1.61,1.73) and FLASH (1.53,1.50) to SB (2.35,2.32) and SB-FLASH (2.28,2.28) for (90º, 270º) in prostate. The difference between peak and valley dose is also clear from the figures, i.e., SB (Figure 4d) and SB-FLASH (Figure 4e) compared to CONV (Figure 4a) and FLASH (Figures 4b). The improved PVDR by SB and SB-FLASH from CONV and FLASH is also demonstrated by BEV dose plane plots in Figure 4g and dose profile plots in Figure 4h. We note that the PVDR values in this work are relatively lower than those in other references because we use the mean values of the peak and valley areas.

*3.2. $P_{FLASH}$*

From Tables 1-3, FLASH and MB-FLASH have very high $P_{FLASH}$, while CONV and MB don't have FLASH effects around the target. We could also find that MB-FLASH has higher $P_{FLASH}$ than FLASH for most fields, i.e., an increase from FLASH (65.0%, 67.4%, 68.4%, 67.5%) to MB-FLASH (70.1%, 74.1%, 71.5%, 76.9%) for (45º, 135º, 225º,315º) in HN, from FLASH (66.1%, 68.6%, 68.8%) to MB-FLASH (73.1%, 73.6%, 73.8%) for (135º, 225º,315º) in abdomen, from FLASH (61.8%, 54.7%) to MB-FLASH (68.3%, 62.8%) for (120º, 240º) in lung and the same 66.1% for 45º in abdomen and a decrease from FLASH (65.6%) to MB-FLASH (63.0%) for 0º in lung. This means that the peak dose area in MB-FLASH plans can help increase the area of FLASH effects, since there is a threshold for the dose to achieve FLASH effects.

From Table 4, FLASH and SB-FLASH have very high $P_{FLASH}$, while CONV and SB don't have FLASH effects around the target. We could also find that SB-FLASH has a little lower $P_{FLASH}$ than FLASH due to separation of the primary beam, i.e., a decrease from FLASH (84.4%, 82.8%) to SB-FLASH (65.7%, 66.8%) for (90º, 270º) in prostate.

*3.3. Physical dose*

The comparison of CI values and CTV1cm mean dose values in Tables 1-4 indicates that both the techniques of heterogeneous dose (for high PVDR purpose) and the dose rate optimization (for high dose rate purpose) may sacrifice the dose coverage. For example, CI has a decrease from CONV (0.79) to FLASH (0.68) and MB (0.67) and MB-FLASH (0.47) in HN, from CONV (0.95) to MB (0.91) and MB-FLASH (0.63) in abdomen, from CONV (0.82) to FLASH (0.73) and MB (0.79) and MB-FLASH (0.70) in lung, from CONV (0.87) to FLASH (0.83) and SB-FLASH (0.81) in prostate; CTV1cm mean dose has an increase from CONV (23.8 Gy) to FLASH (29.8 Gy) and MB (24.8 Gy) and MB-FLASH (32.3 Gy) in HN, from CONV (24.4 Gy) to FLASH (30.4 Gy) and MB (25.0 Gy) and MB-FLASH (35.5 Gy) in abdomen, from CONV (28.4 Gy) to FLASH (30.8 Gy) and MB (28.8 Gy) and MB-FLASH (32.9 Gy) in lung, from CONV (22.9 Gy) to FLASH (24.3 Gy) and SB-FLASH (24.1 Gy) in prostate.

*3.4. Effective dose*

Given the tradeoff between dose rate optimization and dose optimization, the physical dose d and the FLASH dose modifying factor (DMF) were combined into FLASH effective dose $d_{eff}$ [17]. FLASH effective dose parameters calculated for FLASH, MB-FLASH and SB-FLASH in Tables 1-4 in red. The results demonstrate that the dose rate optimization improved the FLASH effective dose from CONV. For example, CI has an increase from CONV (0.79) to FLASH (0.98) and MB-FLASH (0.98) in HN, from CONV (0.95) to FLASH (1.00) and MB-FLASH (1.00) in abdomen, from CONV (0.82) to FLASH (0.99) and MB-FLASH (0.98) in lung, from CONV (0.87) to FLASH (0.98) and SB-FLASH (0.98) in prostate; $V_{30}$ has a decrease from CONV (10.0%) to FLASH (7.9%) and MB-FLASH (9.3%) for lung in lung; $V_{15}$ has a decrease from CONV (38.3%) to FLASH (1.9%) and MB-FLASH (2.7%) for small bowel in abdomen; $D_1$ has a decrease from CONV (34.3 Gy) to FLASH (26.7 Gy) and MB-FLASH (27.6 Gy) for brainstem in HN, $D_{10}$ has a decrease from CONV (27.4 Gy) to FLASH (20.3 Gy) and SB-FLASH (20.2 Gy) for bladder in prostate. FLASH effective dose plots in Figures 1c, 1f, 2c, 2f, 3c, 3f, 4c, 4f show that FLASH, MB-FLASH and SB-FLASH can greatly reduce the high dose area near the target. The DVH plots together with physical dose and FLASH effective dose are shown in Figures 1i, 1j, 2i, 2j, 2k, 3i, 3j, 3k, 4j, 4k, 4l, 4m which demonstrate the advantage of FLASH, MB-FLASH and SB-FLASH in protecting the OARs from high dose areas.

Table 1. Plan parameters for HN.

| Quantity (Unit) | CONV (d) | FLASH (d/$d_{eff}$) | MB (d) | MB-FLASH (d/$d_{eff}$) |
|---|---|---|---|---|
| CI | 0.79 | 0.68/0.98 | 0.67 | 0.47/0.98 |
| $D_{mean,BS}$ (Gy) | 7.8 | 13.7/11.8 | 7.7 | 13.8/11.4 |
| $V_{70,BS}$ (%) | 6.1 | 18.5/0.0 | 6.0 | 26.9/0.6 |
| $D_{5,BS}$ (Gy) | 29.4 | 32.1/25.2 | 29.2 | 34.0/26.1 |
| $D_{1,BS}$ (Gy) | 34.3 | 34.6/26.7 | 35.2 | 36.8/27.6 |
| $D_{mean,CTV1cm}$ (Gy) | 23.8 | 29.8/22.6 | 24.8 | 32.3/24.0 |
| $V_{80,CTV1cm}$ (%) | 28.2 | 43.4/0.0 | 32.1 | 56.2/2.0 |
| $D_{20,CTV1cm}$ (Gy) | 35.2 | 37.9/26.7 | 36.7 | 40.7/28.6 |
| $D_{5,CTV1cm}$ (Gy) | 40.0 | 41.0/28.7 | 41.5 | 43.7/30.7 |
| $PVDR_{4cm,45°}$ | 1.41 | 1.22 | 2.52 | 2.76 |
| $PVDR_{2cm,135°}$ | 1.47 | 1.43 | 2.31 | 2.48 |
| $PVDR_{2cm,225°}$ | 1.31 | 1.58 | 2.76 | 2.88 |
| $PVDR_{4cm,315°}$ | 1.11 | 1.15 | 2.77 | 2.84 |
| $P_{FLASH, 45°}$ (%) | - | 65.0 | - | 70.1 |
| $P_{FLASH, 135°}$ (%) | - | 67.4 | - | 74.1 |
| $P_{FLASH, 225°}$ (%) | - | 68.4 | - | 71.5 |
| $P_{FLASH, 315°}$ (%) | - | 67.5 | - | 76.9 |

Table 2. Plan parameters for abdomen.

| Quantity (Unit) | CONV (d) | FLASH (d/$d_{eff}$) | MB (d) | MB-FLASH (d/$d_{eff}$) |
|---|---|---|---|---|
| CI | 0.95 | 0.97/1.00 | 0.91 | 0.63/1.00 |
| $D_{mean,SC}$ (Gy) | 2.5 | 2.8/2.1 | 2.5 | 2.8/2.2 |
| $V_{10,SC}$ (%) | 35.9 | 38.9/34.1 | 25.5 | 28.4/24.6 |
| $D_{20,SC}$ (Gy) | 6.5 | 6.9/4.8 | 5.1 | 5.5/4.4 |
| $D_{10,SC}$ (Gy) | 6.7 | 7.1/5.0 | 8.5 | 9.0/6.3 |
| $D_{mean,SB}$ (Gy) | 3.1 | 3.5/2.6 | 3.1 | 3.5/2.6 |
| $V_{15,SB}$ (%) | 38.3 | 45.1/1.9 | 34.7 | 39.3/2.7 |
| $D_{20,SB}$ (Gy) | 6.9 | 7.1/5.0 | 6.7 | 7.1/5.1 |
| $D_{5,SB}$ (Gy) | 7.1 | 7.4/5.2 | 7.5 | 8.0/5.8 |
| $D_{mean,CTV1cm}$ (Gy) | 24.4 | 30.4/23.0 | 25.0 | 35.5/26.8 |
| $V_{80,CTV1cm}$ (%) | 27.6 | 46.6/0.0 | 31.1 | 69.0/13.0 |
| $D_{20,CTV1cm}$ (Gy) | 34.6 | 38.4/27.0 | 35.6 | 43.0/30.9 |
| $D_{5,CTV1cm}$ (Gy) | 38.6 | 40.7/28.6 | 39.7 | 46.6/33.7 |
| $PVDR_{2cm,45°}$ | 1.2 | 1.2 | 17.8 | 10.2 |
| $PVDR_{3cm,135°}$ | 1.1 | 1.1 | 2.6 | 2.6 |
| $PVDR_{5.5cm,225°}$ | 1.0 | 1.0 | 4.4 | 4.4 |
| $PVDR_{6cm,315°}$ | 1.0 | 1.1 | 3.4 | 3.5 |
| $P_{FLASH, 45°}$ (%) | - | 66.1 | - | 66.1 |
| $P_{FLASH, 135°}$ (%) | - | 66.1 | - | 73.1 |
| $P_{FLASH, 225°}$ (%) | - | 68.6 | - | 73.6 |
| $P_{FLASH, 315°}$ (%) | - | 68.8 | - | 73.8 |

Table 3. Plan parameters for lung.

| Quantity (Unit) | CONV (d) | FLASH (d/$d_{eff}$) | MB (d) | MB-FLASH (d/$d_{eff}$) |
|---|---|---|---|---|
| CI | 0.82 | 0.73/0.99 | 0.79 | 0.70/0.98 |
| $D_{mean,Lung}$ (Gy) | 3.2 | 3.4/2.7 | 3.2 | 3.6/2.9 |
| $V_{30,Lung}$ (%) | 10.0 | 10.5/7.9 | 10.1 | 11.2/9.3 |
| $D_{20,Lung}$ (Gy) | 5.2 | 5.4/4.2 | 3.9 | 4.6/4.1 |
| $D_{5,Lung}$ (Gy) | 19.5 | 20.8/16.2 | 19.8 | 21.9/17.8 |
| $D_{mean,Eso}$ (Gy) | 2.9 | 3.1/2.6 | 3.0 | 3.2/2.7 |
| $V_{20,Eso}$ (%) | 8.5 | 11.8/5.5 | 12.8 | 14.6/6.3 |
| $D_{20,Eso}$ (Gy) | 6.9 | 7.1/5.8 | 7.2 | 7.4/5.8 |
| $D_{10,Eso}$ (Gy) | 7.8 | 8.2/7.0 | 8.4 | 8.8/7.0 |
| $D_{mean,CTV1cm}$ (Gy) | 28.4 | 30.8/23.5 | 28.8 | 32.9/25.3 |
| $V_{80,CTV1cm}$ (%) | 41.9 | 50.4/3.3 | 44.2 | 60.1/6.7 |
| $D_{20,CTV1cm}$ (Gy) | 38.0 | 40.3/29.0 | 38.7 | 40.8/29.6 |
| $D_{5,CTV1cm}$ (Gy) | 41.5 | 43.6/31.4 | 41.8 | 43.0/32.6 |
| $PVDR_{2cm,0°}$ | 1.3 | 1.4 | 2.2 | 2.2 |
| $PVDR_{5cm,120°}$ | 1.1 | 1.2 | 3.3 | 3.4 |
| $PVDR_{9cm,240°}$ | 1.1 | 1.1 | 6.1 | 6.1 |
| $P_{FLASH, 0°}$ (%) | - | 65.6 | - | 63.0 |
| $P_{FLASH, 120°}$ (%) | - | 61.8 | - | 68.3 |
| $P_{FLASH, 240°}$ (%) | - | 54.7 | - | 62.8 |

Table 4. Plan parameters for prostate.

| Quantity (Unit) | CONV (d) | FLASH (d/$d_{eff}$) | SB (d) | SB-FLASH (d/$d_{eff}$) |
|---|---|---|---|---|
| CI | 0.87 | 0.83/0.98 | 0.87 | 0.81/0.98 |
| $D_{mean,Bladder}$ (Gy) | 7.9 | 9.2/7.0 | 7.8 | 8.9/6.9 |
| $V_{60, Bladder}$ (%) | 19.4 | 23.5/12.9 | 18.9 | 22.6/12.5 |
| $D_{10, Bladder}$ (Gy) | 27.4 | 28.8/20.3 | 27.3 | 28.7/20.2 |
| $D_{5, Bladder}$ (Gy) | 30.6 | 31.2/21.8 | 30.5 | 31.2/21.9 |
| $D_{mean,Rectum}$ (Gy) | 9.0 | 9.9/7.3 | 8.9 | 9.6/7.3 |
| $V_{60, Rectum}$ (%) | 23.6 | 26.4/14.9 | 23.5 | 24.9/14.5 |
| $D_{10, Rectum}$ (Gy) | 28.8 | 29.6/20.7 | 28.8 | 29.6/20.7 |
| $D_{5, Rectum}$ (Gy) | 30.9 | 31.3/21.9 | 31.0 | 31.3/22.0 |
| $D_{mean,CTV1cm}$ (Gy) | 22.9 | 24.3/17.6 | 22.8 | 24.1/17.8 |
| $V_{80,CTV1cm}$ (%) | 44.7 | 50.0/0.0 | 43.9 | 48.8/0.0 |
| $D_{20,CTV1cm}$ (Gy) | 30.1 | 30.7/21.6 | 30.1 | 30.7/21.6 |
| $D_{5,CTV1cm}$ (Gy) | 31.9 | 32.1/22.5 | 31.9 | 32.2/22.7 |
| $PVDR_{7cm,90°}$ | 1.61 | 1.53 | 2.35 | 2.28 |
| $PVDR_{7cm,270°}$ | 1.73 | 1.50 | 2.32 | 2.28 |
| $P_{FLASH, 90°}$ (%) | - | 84.4 | | 65.7 |
| $P_{FLASH, 270°}$ (%) | - | 82.8 | | 66.8 |

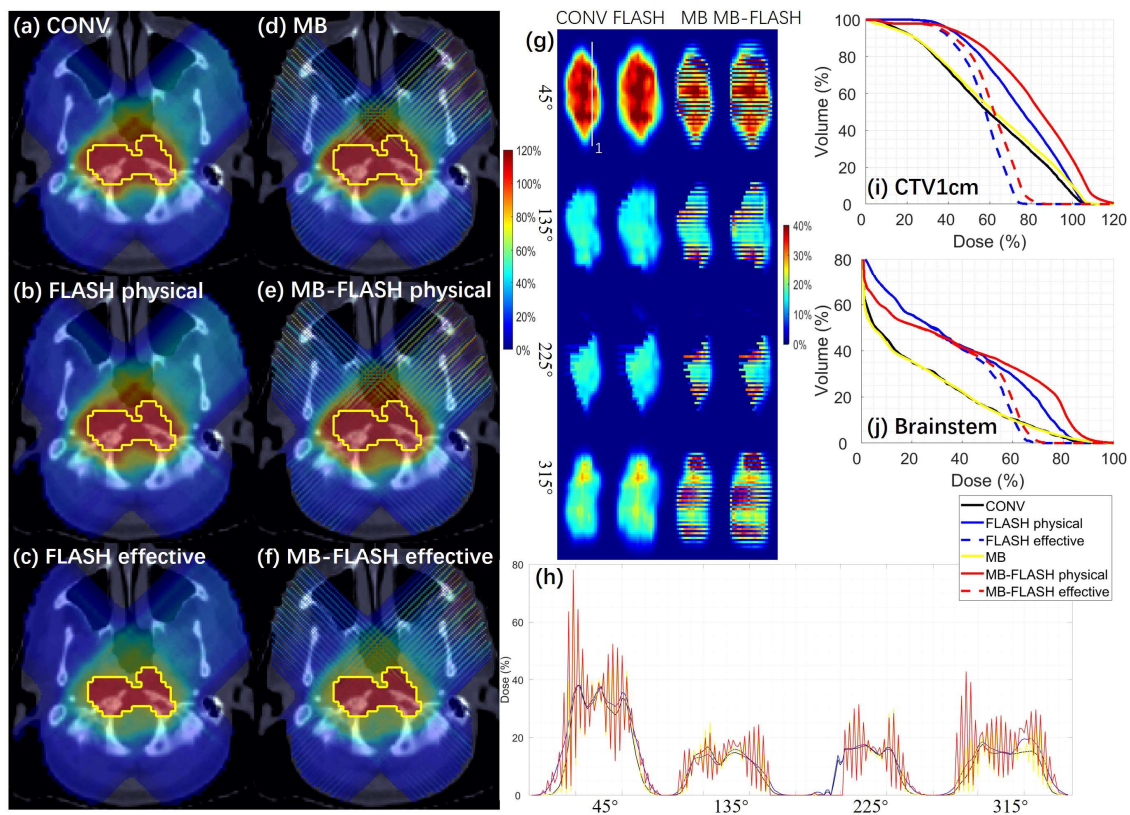

Figure 1. HN. (a) CONV dose plot. (b) FLASH physical dose plot. (c) FLASH effective dose plot. (d) MB dose plot. (e) MB-FLASH physical dose plot. (f) MB-FLASH effective dose plot. (g) plots of BEV dose slices at 4cm depth from 45° beam, 2cm depth from 135° beam, 2cm depth from 225° beam, 4cm depth from 315° beam. (h) plots of BEV dose profiles at position 1 in (g). (i) DVH for CTV1cm, a 1cm expansion ring of CTV. (j) DVH for Brainstem. The dose plot window is [0%, 120%] and CTV is highlighted in dose plots.

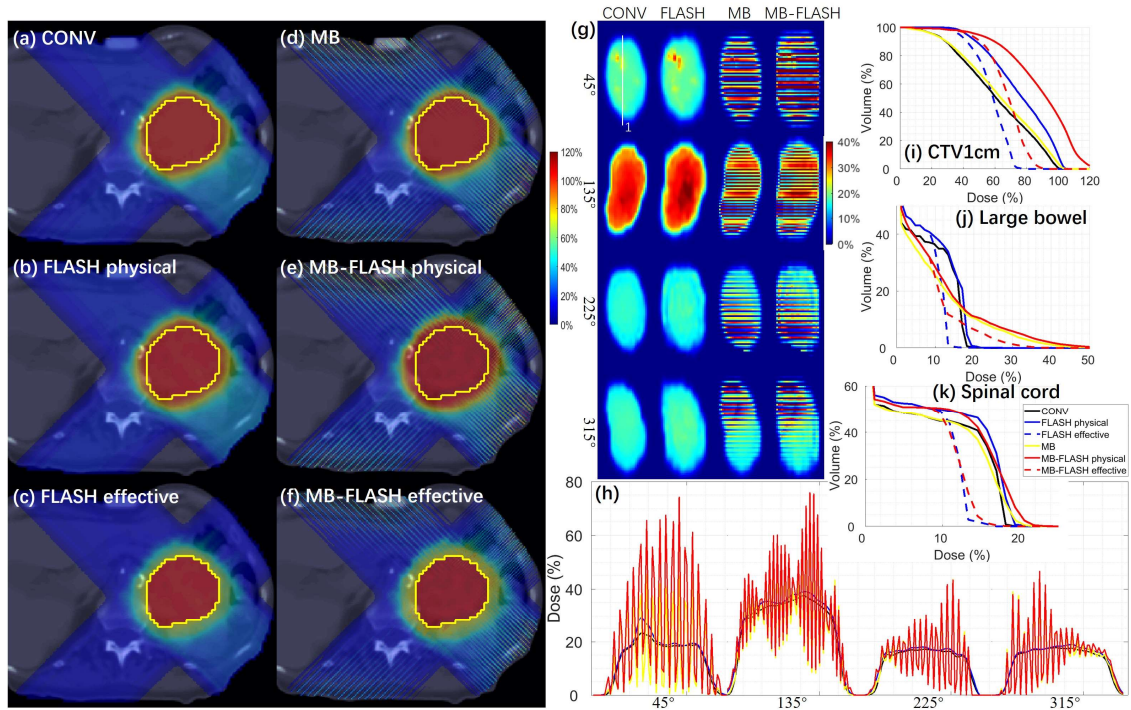

Figure 2. Abdomen. (a) CONV dose plot. (b) FLASH physical dose plot. (c) FLASH effective dose plot. (d) MB dose plot. (e) MB-FLASH physical dose plot. (f) MB-FLASH effective dose plot. (g) plots of BEV dose slices at 2cm depth from 45° beam, 3cm depth from 135° beam, 5.5cm depth from 225° beam, 6cm depth from 315° beam. (h) plots of BEV dose profiles at position 1 in (g). (i) DVH for CTV1cm, a 1cm expansion ring of CTV. (j) DVH for large bowel. (k) DVH for spinal cord. The dose plot window is [0%, 120%] and CTV is highlighted in dose plots.

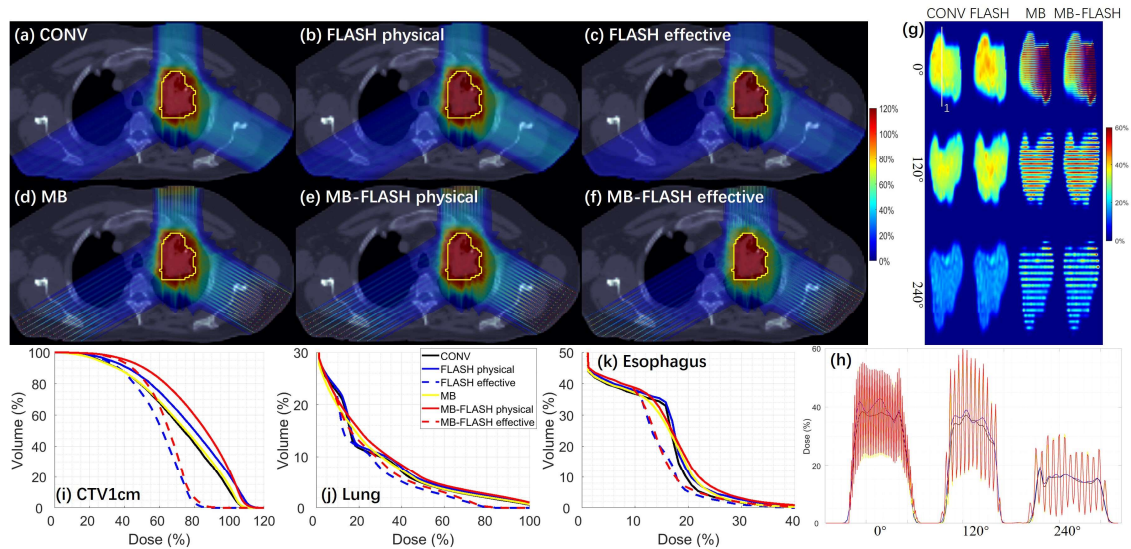

Figure 3. Lung. (a) CONV dose plot. (b) FLASH physical dose plot. (c) FLASH effective dose plot. (d) MB dose plot. (e) MB-FLASH physical dose plot. (f) MB-FLASH effective dose plot. (g) plots of BEV dose slices at 2cm depth from 0° beam, 5cm depth from 120° beam, 9cm depth from 240° beam. (h) plots of BEV dose profiles at position 1 in (g). (i) DVH for CTV1cm, a 1cm expansion ring of CTV. (j) DVH for lung. (k) DVH for esophagus. The dose plot window is [0%, 120%] and CTV is highlighted in dose plots.

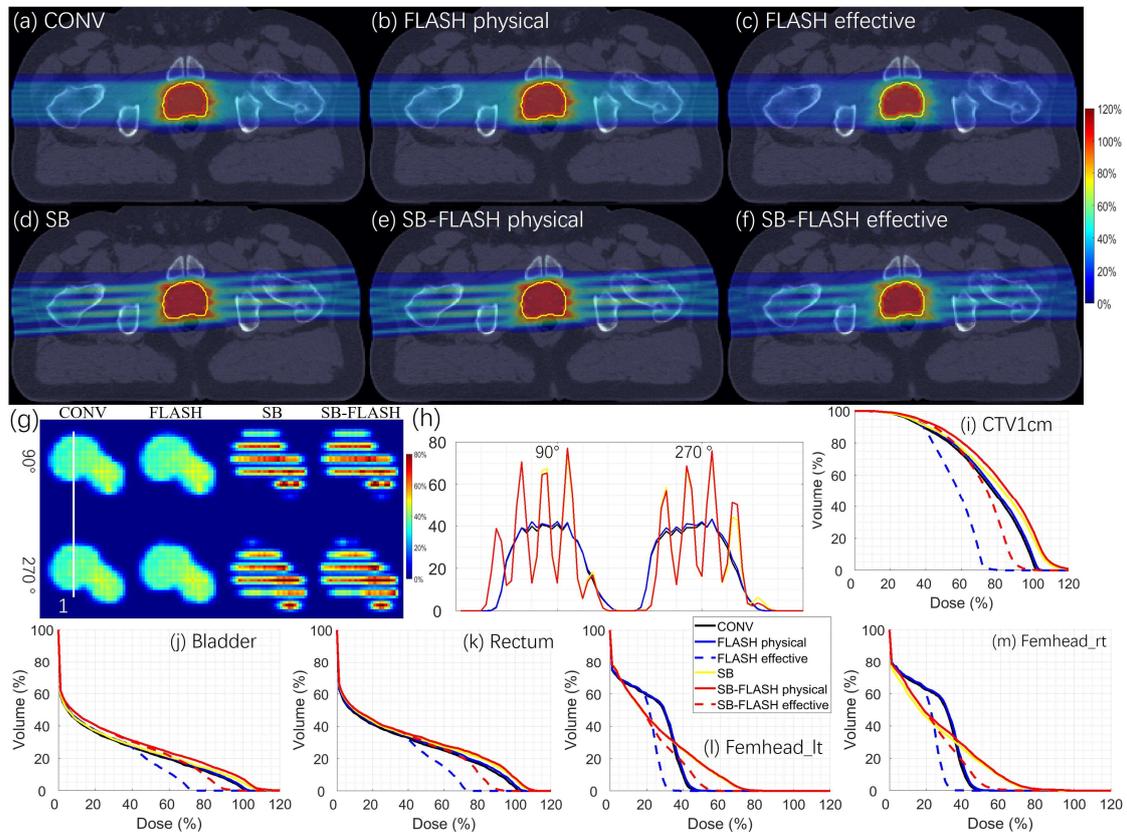

Figure 4. Prostate. (a) CONV dose plot. (b) FLASH physical dose plot. (c) FLASH effective dose plot. (d) SB dose plot. (e) SB-FLASH physical dose plot. (f) SB-FLASH effective dose plot. (g) plots of BEV dose slices at 7cm depth from 90° beam, 7cm depth from 270° beam. (h) plots of BEV dose profiles at position 1 in (g). (i) DVH for CTV1cm, a 1cm expansion ring of CTV. (j) DVH for bladder. (k) DVH for rectum. (l) DVH for left femhead. (m) DVH for right femhead. The dose plot window is [0%, 120%] and CTV is highlighted in dose plots.

## 4. Discussion

This study investigates the combination of FLASH-RT and SFRT in proton treatment planning, aiming to maximize normal tissue sparing by harnessing the unique advantages of each technique. The high PVDR and extensive FLASH effect coverage demonstrated in our results suggest that this combined approach can effectively protect normal tissue, utilizing the FLASH effect at greater depths and high PVDR at shallow-to-intermediate depths. Both evaluated modalities (MB-FLASH and SB-FLASH) achieve a uniform CTV dose and spatially fractionated dose in the normal tissue. The delivery time was reduced by using the highest-energy beam and dose rate constraints have been applied to generate sufficient dose rate (>40 Gy/s) coverage. These dosimetric advantages may translate to better clinical outcomes.

As demonstrated in the treatment plan comparisons in Section 3, SFRT-FLASH proves highly effective. Imposing a high-dose-rate constraint for the ROI did not negatively impact PVDR in shallow-to-intermediate depths, and the spatially fractionated dose distribution at these depths may increase, or minimally affect, the dose rate in the ROI. Comparisons across various biological parameters indicate that our proposed method offers enhanced normal tissue protection at different depths relative to conventional treatment plans.

For the proton GRID modality, we employed the scissor beam technique [38], avoiding the need for a physical collimator, which reduces the MMU requirement. We set MMU as around 100 in SB-FLASH plans to balance plan quality with sufficient FLASH effect coverage. For MB-FLASH plans, we used a multi-slit collimator, which can block radiation with large ctc distances and increase the MMU requirement to approximately 1000. In future studies, it may be worth considering alternative methods that do not require collimators [48] to generate minibeam dose patterns, which could enhance the integration of MBRT and FLASH-RT.

In this study, we took into account both dose and dose rate thresholds to accurately estimate FLASH effect coverage. However, in our optimization approach, we focused on increasing the dose rate in the ROI by applying dose rate constraints, without additional constraints to push the dose in the ROI to the FLASH threshold level. With a hypofractionated prescription dose, the FLASH, MB-FLASH, and SB-FLASH plans achieved substantial FLASH effect coverage (5 Gy dose threshold) within the ROI, even without enforcing additional dose constraints. Incorporating dose constraints aimed at achieving the threshold dose may compromise plan quality by increasing the overall dose. With a clearer understanding of the FLASH mechanism in the future, new optimization methods may be developed to jointly consider the threshold of dose and dose rate.

## 5. Conclusion

We have developed a novel proton treatment planning approach that combines FLASH-RT and SFRT to enhance normal tissue protection. The proposed method, with its two modalities. Two different modalities, MB-FLASH and SB-FLASH, achieve both high PVDR at shallow-to-intermediate depths and significant FLASH effect coverage near the target, enabling enhanced protection for the normal tissue.